\RequirePackage{ifpdf}

\documentclass[a4paper,11pt]{article}
\pdfoutput=1 

\usepackage{jheppub} 

\usepackage[T1]{fontenc} 

\usepackage{ifpdf} 

\usepackage{flexisym}
\usepackage{breqn}
\usepackage{comment}

\usepackage{epsf,epsfig,epstopdf}
\usepackage{graphics}
\usepackage{graphicx}
\usepackage[usenames,dvipsnames]{xcolor}
\definecolor{darkgreen}{RGB}{34,139,34}

\usepackage{enumerate}

\usepackage{url}
\usepackage{amsmath,amsfonts,amssymb}

\usepackage[utf8]{inputenc}

\newcommand{\kur}[1]{\mathcal{#1}}
\newcommand{\id}{\mathrm{d}}
\newcommand{\nn}{\nonumber}
\newcommand{\define}{\equiv}
\newcommand{\prop}{\varpropto}
\newcommand{\Li}{\text{Li}}
\newcommand{\eps}{\epsilon}

\author{Hjalte Frellesvig and Costas G. Papadopoulos}

\affiliation{Institute of Nuclear and Particle Physics, NCSR `Demokritos', Agia Paraskevi, 15310, Greece}
\emailAdd{frellesvig@inp.demokritos.gr}
\emailAdd{costas.papadopoulos@cern.ch}


\title{Cuts of Feynman Integrals in Baikov representation}

\abstract{
Based on the Baikov representation, we present a systematic approach to compute cuts of Feynman Integrals, appropriately defined in $d$ dimensions.
The information provided by these computations may be used to determine the class of functions needed to
analytically express the full integrals.  
}

\ifpdf
\fi

\begin{document}
\unitlength1cm
\maketitle
\allowdisplaybreaks

\section{Introduction}
\label{sIntro}


It is almost seventy years from the time Feynman Integrals (FI) were first introduced~\cite{Feynman:1949zx,Dyson:1949bp,Dyson:1949ha} and forty-five years 
since the dimensional regularisation~\cite{tHooft:1972tcz} set up the framework for an efficient use of loop integrals 
in computing scattering matrix elements, and still the frontier of multi-scale multi-loop integral calculations (maximal both in number of scales and number of loops) is determined by the planar five-point two-loop on-shell massless integrals~\cite{Gehrmann:2015bfy,Papadopoulos:2015jft}, 
recently computed\footnote{Complete results, including physical region kinematics, are presented in~\cite{Papadopoulos:2015jft}. Notice that numerical codes, 
like for instance {\tt SecDec}~\cite{Borowka:2015mxa}, can reproduce analytic results only at Euclidean region kinematics; results for physical region kinematics are not supported due to poor numerical convergence.}.  
On the other hand, in order to keep up with the increasing experimental accuracy as more data is collected at the LHC, more precise theoretical predictions and higher loop calculations are required~\cite{Andersen:2014efa}.

In the last years our understanding of the reduction of one-loop amplitudes to a set of Master Integrals (MI), a minimal set of FI that form a basis, 
either based on unitarity methods~\cite{Bern:1994cg,Bern:1994zx,Berger:2008sj} 
or at the integrand level via the OPP method~\cite{Ossola:2006us,Ossola:2008xq}, has drastically changed the way one-loop calculations are preformed resulting in many fully automated numerical tools (some reviews on the topic are~\cite{AlcarazMaestre:2012vp,Ellis:2011cr,vanDeurzen:2015jmn}), making the next-to-leading order (NLO) approximation the default precision for theoretical predictions at the LHC.
In the recent years, progress has been made also towards the extension of these reduction methods for two-loop amplitudes at the integral~\cite{Gluza:2010ws,Kosower:2011ty,CaronHuot:2012ab,Johansson:2012zv,Johansson:2012sf,Johansson:2013sda,Sogaard:2013fpa,Ita:2015tya,Johansson:2015ava,Mastrolia:2016dhn} 
as well as the integrand~\cite{Mastrolia:2011pr,Badger:2012dp,Mastrolia:2012wf,Badger:2013gxa,Papadopoulos:2013hra,Badger:2015lda} level. Two-loop MI are defined 
using the {\it integration by parts} (IBP) identities~\cite{Chetyrkin:1981qh,Tkachov:1981wb,Laporta:2001dd}, an indispensable tool beyond one loop.
Contrary to the one-loop case, where MI have been known for a long time already~\cite{'tHooft:1978xw}, a complete library of MI at two-loops is still missing.
At the moment this is the main obstacle to obtain a fully automated NNLO calculation framework similar to the one-loop one, that will satisfy the precision requirements at the LHC~\cite{Andersen:2014efa}.

Many methods have been introduced in order to compute MI~\cite{Smirnov:2012gma}. 
The overall most successful one, is based on expressing the FI in terms of an integral representation over Feynman parameters, involving the two well-known Symanzik Polynomials $U$ and $F$~\cite{Bogner:2010kv}.
The introduction of the sector decomposition~\cite{Hepp:1966eg,Roth:1996pd,Binoth:2000ps,Binoth:2003ak,Bogner:2007cr} method resulted in a powerful computational framework for the numerical evaluation of FI, 
see for instance {\tt SecDec}~\cite{Borowka:2015mxa}. An alternative is based on Mellin-Barnes representation~\cite{Smirnov:1999gc,Tausk:1999vh}, implemented in~\cite{Czakon:2005rk}\footnote{See also https://mbtools.hepforge.org}. Nevertheless, the most successful method to calculate multi-scale multi-loop FI is, for the time being,  the differential equations (DE) 
approach~\cite{Kotikov:1990kg,Kotikov:1991pm,Bern:1992em,Remiddi:1997ny,Gehrmann:1999as}, which has been used in the past two decades to calculate various MI at two-loops and beyond. 
Following the work of refs.~\cite{Goncharov:1998kja,Remiddi:1999ew,Goncharov:2001iea}, there has been a building consensus that the so-called {\it Goncharov Polylogarithms} (GPs) form a functional basis for many MI. 
The so-called canonical form of DE, introduced by Henn~\cite{Henn:2013pwa}, manifestly results in MI expressed in terms of GPs~\footnote{For an alternative method in the single scale case see also ref.~\cite{Ablinger:2015tua}}. 
Nevertheless the reduction of a given DE to a canonical form is by no means
fully understood. First of all, despite recent efforts~\cite{Lee:2014ioa,Meyer:2016slj,Gituliar:2017vzm}, 
and the existence of {\it sufficient} conditions that a given MI can be expressed in terms of GPs, no criterion, with practical applicability, that is at the same time {\it necessary and sufficient} 
has been introduced so far. Moreover, it is well known that when for instance enough internal masses
are introduced, MI are not anymore expressible in terms of GPs, and in fact a new class of functions involving elliptic integrals is needed~\cite{Adams:2015gva,Bonciani:2016qxi}. 

In this paper we are studying another representation of FI introduced by Baikov~\cite{Baikov:1996iu,Baikov:1996rk,Smirnov:2003kc,Lee:2010wea,Grozin:2011mt,Lee:2013hzt}. As we will see in Section \ref{sBaikov}, 
where we present a review of its derivation,
it has several nice features, including its conceptual simplicity, a direct factorisation of kinematics and loop `topology', incorporation of IBP identities in a straightforward manner.
We also present, to the best of our knowledge for the first time, a consistent definition of the integration limits which will be important for the computation of cut integrals in $d$ dimensions. 
In Section \ref{sLBL} we elaborate on the loop-by-loop approach within the Baikov representation,  that has a minimal number of integration variables for a given FI. 
In Section \ref{sDE} we present a novel approach to 
obtain DE from the Baikov representation. In Section \ref{sCUT} we introduce the definition of the cut integral in Baikov representation, that satisfies the same DE and the same IBP identities as the uncut one~\cite{Henn:2014qga}, 
and we conjecture that computing the corresponding maximally cut integral~\cite{Primo:2016ebd}
 we may have a necessary and sufficient condition for the expression of the uncut integral in terms of GPs and when 
applied to the whole family of MI on the possibility to obtain a canonical form.  Finally in Appendix \ref{app:derivation} we present an alternative derivation of the Baikov representation for one- and two-loop FI and in
Appendix \ref{app:examples} we collect several examples of maximally cut integrals in Baikov representation that support our findings.

\section{The Baikov representation}
\label{sBaikov}

In this section we introduce the Baikov representation following ref.~\cite{Lee:2010wea,Grozin:2011mt}. An $L$-loop Feynman Integral with $E+1$ external lines can be written in the form
\begin{equation}
{F_{{\alpha_1}...{\alpha_N}}} = \int {\left( {\prod\limits_{i = 1}^L {\frac{{{d^d}{k_i}}}{{i{\pi ^{d/2}}}}} } \right)} \frac{1}{{D_1^{{\alpha_1}}...D_N^{{\alpha_N}}}}
\label{FI}
\end{equation}
with $N = \frac{{L\left( {L + 1} \right)}}{2} + LE$, $\alpha_i$ arbitrary integers, and $D_a$, $a=1,...,N$, inverse Feynman propagators, $P^2-M^2$, where $P$ represents, collectively,  a linear combination of loop and external momenta and $M$ internal masses, as dictated by the Feynman Integral in consideration.

To be more specific, let us define $q_i=k_i, (i=1,...,L)$ the loop momenta and $q_{L+i}=p_i, (i=1,...,E)$, the independent external momenta, $M=L+E$ and $s_{ij}=q_i\cdot q_j$.
Then 
\begin{equation}
{D_a} = \sum\limits_{i = 1}^L {\sum\limits_{j = i}^M {A_{_a}^{ij}{s_{ij}} + {f_a}} }  = \sum\limits_{i = 1}^L {\sum\limits_{j = i}^L {A_{_a}^{ij}{k_i} \cdot {k_j}} }  + \sum\limits_{i = 1}^L {\sum\limits_{j = L+1}^M {A_{_a}^{ij}{k_i} \cdot {p_{j-L}}} }  + {f_a},\,\,\,\,\,\,a = 1, \ldots ,N
\label{eq:x-def}
\end{equation}
where $f_a$ depend on external kinematics and internal masses. $A_{_a}^{ij}$ can be understood as an $N\times N$ matrix, with $a$ running obviously from $1$ to $N$ and with $(ij)$ taking also $N$ values as $i=1,\ldots,L$ and $j=i,\ldots,M$.
The elements of the matrix $A_{_a}^{ij}$ are integer numbers taken  from the set $\left\{-2,-1,0,+1,+2 \right\}$. This matrix is characteristic of the corresponding Feynman graph and can, in a loose sense, 
be associated with the `topology' of the graph. 
Then, by projecting each of the loop momenta $q_i=k_i, (i=1,...,L)$ with respect to the space spanned by the external momenta involved plus a transverse component (for details see~\cite{Grozin:2011mt}), we may write
\begin{equation}
{F_{{\alpha_1}...{\alpha_N}}} = C_{N}^L{\left( {G\left( {{p_1},...,{p_E}} \right)} \right)^{\left( { - d + E + 1} \right)/2}}\int {\frac{{d{x_1}...d{x_N}}}{{x_1^{{\alpha_1}}...\,\,x_N^{{\alpha_N}}}}} P_N^L{\left( {{x_1} - {f_1},...,{x_N} - {f_N}} \right)^{\left( {d - M - 1} \right)/2}}
\label{BR}
\end{equation}
with 
\begin{equation}
C_{N}^L = \frac{{{\pi ^{ - L\left( {L - 1} \right)/4 - LE/2}}}}{{\prod\nolimits_{i = 1}^L {\Gamma \left( {\frac{{d - M + i}}{2}} \right)} }}\det \left( {A_{ij}^a} \right)
\end{equation}
and 
\[P_N^L\left( {{x_1},{x_2},...,{x_N}} \right) = { { {{{ {G\left( {{k_1},...,{k_L},{p_1},...,{p_E}} \right)} }}} } \Big|_{{s_{ij}} = \sum\limits_{a = 1}^N {A_{ij}^a{x_a}\,\,\& \,\,\,{s_{ji}} = {s_{ij}}} }}\]
with $G$ representing the Gram determinant, $G\left( {{q_1}, \ldots ,{q_n}} \right) = \det \left( {{q_i} \cdot {q_j}} \right)$ and $A_{ij}^a$ is the inverse of the topology matrix $A_{_a}^{ij}$. An alternative derivation of the Baikov representation for one- and two-loop FI is given in Appendix \ref{app:derivation}.

Integration-by-parts identities can easily be accommodated in the Baikov representation. The generators of the IBP identities can be cast into the form
\begin{equation}
O_{ij}P_N^L=0
\end{equation}
with the operators $O_{ij}$ given by ($i=1,\ldots,L$)
\begin{equation}
j \le L\left( {{q_j} = {k_j}} \right)\,\,\,\,\,\,\,\,\,\,\,\,\,{O_{ij}} = d{\delta _{ij}} + \sum\limits_{a = 1}^N {\sum\limits_{b = 1}^N {\sum\limits_{m = 1}^M {A_a^{mi}A_{mj}^b\left( {1 + {\delta _{mi}}} \right)\left( {{x_b} - {f_b}} \right)} } } \frac{\partial }{{\partial {x_a}}}
\end{equation}
and
\begin{equation}
j > L\left( {{q_j} = {p_{j - L}}} \right)\,\,\,\,\,\,\,\,{O_{ij}} = \sum\limits_{a = 1}^N {\left( {\sum\limits_{m = 1}^L {\sum\limits_{b = 1}^N {A_a^{mi}A_{mj}^b\left( {1 + {\delta _{mi}}} \right)\left( {{x_b} - {f_b}} \right) + \sum\limits_{m = L + 1}^M {A_a^{mi}{s_{mj}}} } } } \right)} \frac{\partial }{{\partial {x_a}}}
\end{equation}
The derivation of the Baikov representation can easily be implemented in a computer algebra code\footnote{A {\tt Mathematica} script, {\tt Baikov.m}, is provided as an attachment}. 

We conclude this section by elaborating on the limits of the $x_a-$integrations in Eq. (\ref{BR}). 
In order to simplify the discussion, let us start with a generic one-loop configuration defined by 
\[{x_1} =   {k^2} - m_1^2\,, \quad {x_2} = \left( {k + {p_1}} \right)^2 - m_2^2\,, \quad \ldots\,, \quad {x_N} = \left( {k + {p_1} + \ldots + {p_{N - 1}}} \right)^2 - m_N^2\]
Then consider the generic integral ($\alpha_i \ge 0$),
\begin{equation}
F_{\alpha_1 \cdots \alpha_N} = C^1_N G \! \left( p_1, \ldots, p_{N-1} \right)^{(N-d)/2} \int {\frac{{d{x_1}...d{x_N}}}{{x_1^{{\alpha_1}}...\,\,x_N^{{\alpha_N}}}}} {P^1_N}^{\left( {d - N - 1} \right)/2}
\end{equation}
\begin{equation}
C^1_N = \frac{{{\pi ^{ - (N-1)/2}}}}{{{\Gamma \left( {\frac{{d - N + 1}}{2}} \right)} }}\left(\frac{1}{2}\right)^{N-1}
\end{equation}
It is easy to verify that ${P^1_N}$ is a polynomial that is quadratic in the variables $x_a$~\cite{Smirnov:2003kc}, and that obviously when $\alpha_N=0$, the external momentum $p_{N-1}$ decouples, so that
\begin{eqnarray}
{F_{{\alpha_1}...{\alpha_{N-1}}{0}}}& =& {C^1_N} G \! \left( p_1, \ldots, p_{N-1} \right)^{(N-d)/2} \int {\frac{{d{x_1}...d{x_{N - 1}}}}{{x_1^{{\alpha _1}}...\,\,x_{N - 1}^{{\alpha _{N - 1}}}}}\int\limits_{x_N^ - }^{x_N^ + } {d{x_N}} } {P^1_N}^{\left( {d - N - 1} \right)/2}
\nn \\
&=& C^1_{N-1} G \! \left( p_1, \ldots, p_{N-2} \right)^{(N-1-d)/2}  \int {\frac{{d{x_1}...d{x_{N-1}}}}{{x_1^{{\alpha_1}}...\,\,x_{N-1}^{{\alpha_{N-1}}}}}} {P^1_{N-1}}^{\left( {d - \left(N-1\right) - 1} \right)/2}
\nn\\
\end{eqnarray}
where ${P^1_N}\left( {x_N^ + } \right) = {P^1_N}\left( {x_N^ - } \right) = 0$ and 
\begin{eqnarray}
\int\limits_{x_N^ - }^{x_N^ + } {d{x_N}{P^1_N}^{\left( {d - N - 1} \right)/2}}  &=& \frac{{2{\pi ^{1/2}}\Gamma \left( {\frac{{d - N + 1}}{2}} \right)}}{{\Gamma \left( {\frac{{d - N + 2}}{2}} \right)}}
G \left( p_1, \ldots, p_{N-1} \right)^{(d-N)/2}
G \left( p_1, \ldots, p_{N-2} \right)^{(N-1-d)/2}
\nn\\
&\times &
{P^1_{N - 1}}^{\left( {d - \left( {N - 1} \right) - 1} \right)/2}
\nn
\end{eqnarray}
using $P_N^1 = \frac{1}{4}G\left( {{p_1}, \ldots ,{p_{N - 2}}} \right)\left( {x_N^ +  - {x_N}} \right)\left( {{x_N} - x_N^ - } \right)$ and ${\left( {x_N^ +  - x_N^ - } \right)^2} = 16\frac{{G\left( {{p_1}, \ldots ,{p_{N - 1}}} \right)}}{{G{{\left( {{p_1}, \ldots ,{p_{N - 2}}} \right)}^2}}}P_{N - 1}^1$.
This can be repeated straightforwardly for all variables except $x_1=k^2-m_1^2$ whose integration limits are simply derived from the  $k-$modulus integration limits. The generalisation to the two-loop case is straightforward, with the integration at each step
performed over the $x-$variables involving a given external momentum, and the last ones derived by the corresponding $k_1-$ and $k_2-$modulus integration limits. We have checked both analytically and numerically that
the limits, as defined above, reproduce  the known results for several examples at one and two loops.

\section{The loop-by-loop approach}
\label{sLBL}

\begin{figure}[t!]
\centering
\includegraphics[width=0.35 \linewidth]{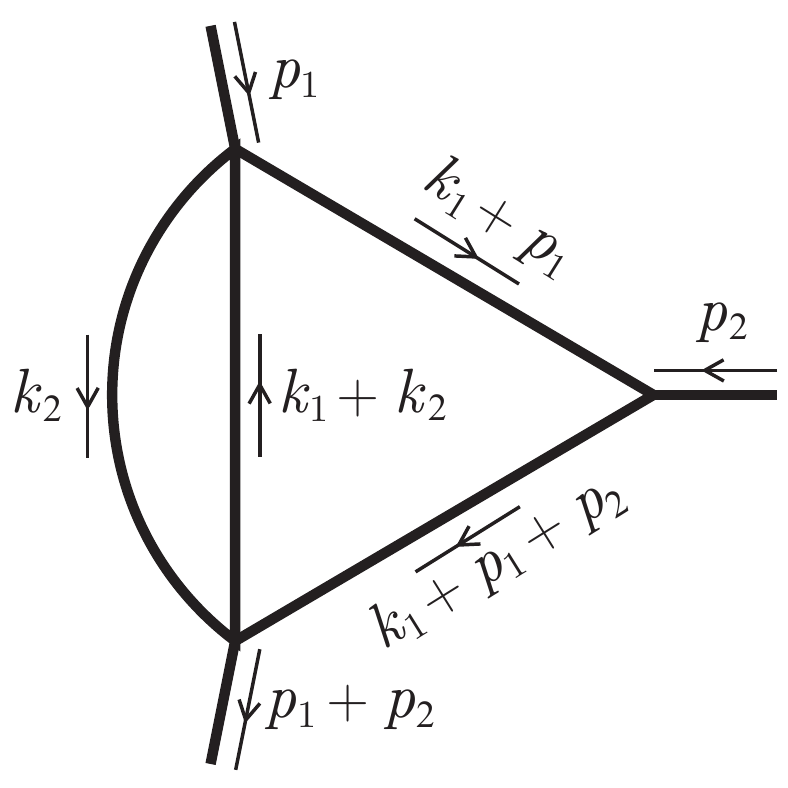}
\caption{The two-loop three-point graph considered in the text.}
  \label{fig:2L3P}
\end{figure}

At one loop the number of Baikov variables and the number of propagators of a generic Feynman Integral can be the same. At higher loops this is not the case anymore. Consider for instance the double-box, with
$E=3$ independent external momenta and $L=2$, giving $N=9$ and $M=5$. In that case we have to integrate over nine variables whereas the scalar double-box graph has only seven denominators. 
Of course this mismatch is the origin of the well known irreducible scalar products (ISP) appearing for $L\ge 2$. It is eventually desirable to have a representation with the minimal number of integration variables.
This can be achieved by considering the projection over the space spanned by the external momenta of each loop integration momentum separately. As an example to illustrate the concept, consider the
two-loop three-point Feynman Integral of Figure~\ref{fig:2L3P}. The standard Baikov representation is based on the following set of seven inverse propagators
\begin{align}
x_1 &= k_1^2-m_1^2 & x_2 &= (k_1+p_1)^2 - m_2^2 & x_3 &= (k_1+p_1+p_2)^2-m_3^2 \nn \\
x_4 &= k_2^2-m_4^2 & x_5 &= (k_2-p_1)^2 - m_5^2 & x_6 &= (k_2-p_1-p_2)^2-m_6^2 \\
x_7 &= (k_1+k_2)^2 - m_7^2 \!\!\!\!\!\!\!\!\!\!\!\!\!\!\!\!\!\!\!\! \nn
\end{align}

\begin{equation}
{F_{0111001}} = C_3^2{\left( {G\left( {{p_1},{p_2}} \right)} \right)^{ - \left( {d - 3} \right)/2}}\int {\frac{{d{x_1} \ldots d{x_7}}}{{{x_2}{x_3}{x_4}{x_7}}}{{\left( {P_3^2} \right)}^{\left( {d - 5} \right)/2}}} 
\label{2L3P-g}
\end{equation}

For the integral ${F_{0111001}}$, in the loop-by-loop approach one can consider the $k_1$ and $k_2$ integrations separately. 
Starting with $k_2$ integration it is easy to see that we have a two-point function with external momentum $k_1$.
\begin{equation}
{F_{0111001}} = \int {\frac{{{d^d}{k_1}}}{{i{\pi ^{d/2}}}}\frac{{{d^d}{k_2}}}{{i{\pi ^{d/2}}}}\frac{1}{{{x_2}{x_3}{x_4}{x_7}}}}  = C_2^1\int {\frac{{{d^d}{k_1}}}{{i{\pi ^{d/2}}}}\frac{1}{{{x_2}{x_3}}}{{\left( {{x_1} + m_1^2} \right)}^{ - \left( {d - 2} \right)/2}}\int {\frac{{d{x_4}d{x_7}}}{{{x_4}{x_7}}}} {{\left( {P_2^1} \right)}^{\left( {d - 3} \right)/2}}} 
\end{equation}
Then for the $k_1$ integration we have a three-point integral with 
two external momenta $p_1$ and $p_2$. 
\begin{equation}
{F_{0111001}} = C_2^1C_3^1{\left( {G\left( {{p_1},{p_2}} \right)} \right)^{ - \left( {d - 3} \right)/2}}\int {\frac{{d{x_1}d{x_2}d{x_3}d{x_4}d{x_7}}}{{{x_2}{x_3}{x_4}{x_7}}}{{\left( {{x_1} + m_1^2} \right)}^{ - \left( {d - 2} \right)/2}}{{\left( {P_3^1} \right)}^{\left( {d - 5} \right)/2}}} {\left( {P_2^1} \right)^{\left( {d - 3} \right)/2}}
\label{2L3P-ll}
\end{equation}
Notice that the same result can be obtained from Eq.~(\ref{2L3P-g}) integrating out $x_5$ and $x_6$, as detailed at the end of the previous section. Nevertheless, the loop-by-loop approach offers an alternative
way of obtaining a minimal number of integration variables.

\section{Deriving differential equations}
\label{sDE}
It is instructive to study the way differential equations with respect to external kinematics and masses can be obtained in Baikov representation. 
Differential equations are usually written in terms of external kinematical invariants, $s_{ij}=(p_i+p_j)^2$ and internal masses, $m_i^2$. 
In the standard approach, since the integral in the momentum-space representation is not an explicit function of the kinematical invariants, derivatives with respect to external momenta, $p_j^\mu \frac{\partial }{{\partial p_i^\mu }}$, are used.
In Baikov representation though, the dependence on external kinematical invariants and internal masses is explicit.
Indeed in Eq.~(\ref{BR}), it is easy to identify two terms that depends
on the external kinematics and/or masses, namely the overall factor $G \left( p_1, \ldots , p_E \right)^{\left( { - d + E + 1} \right)/2}$ and the Baikov polynomial $P_N^L$ itself. The differentiation of the first
factor causes no problem since the result is expressed in terms of the original integral. For the Baikov polynomial this is not so, since the derivative introduces a different integrand that is not directly expressible
in terms of FI, Eq.~(\ref{FI}).  To be more specific, let us denote by $X$ a generic kinematical variable, for instance a Mandelstam invariant $X=\left(p_i+p_j\right)^2$ or an internal mass $X=m_i^2$. Then
\begin{eqnarray}
\frac{\partial }{{\partial X}}{F_{{\alpha _1}...{\alpha _N}}} &=& \left( {\frac{{ - d + E + 1}}{2}} \right)\left(\frac{1}{G}\frac{{\partial G}}{{\partial X}}\right){F_{{\alpha _1}...{\alpha _N}}}   \\
&+&C_N^L{G^{\left( { - d + E + 1} \right)/2}}\int {\frac{{d{x_1}...d{x_N}}}{{x_1^{{\alpha _1}}...\,\,x_N^{{\alpha _N}}}}} {P_N^L}^{\left( {d - M - 1} \right)/2} \left[ {\left( {\frac{{d - M - 1}}{2}} \right)\frac{1}{{P_N^L}}\frac{{\partial P_N^L}}{{\partial X}}} \right] \nonumber 
\end{eqnarray}
where  $G$ is used for ${G\left( {{p_1},...,{p_E}} \right)} $. Based on the fact that the derivatives 
$\frac{{\partial P_N^L}}{{\partial X}},\frac{{\partial P_N^L}}{{\partial {x_a}}}$ are polynomials in $x_a$, 
the idea is to turn the derivative with respect to $X$ into derivatives with respect to $x_a$. This can be achieved by the equation, known as the syzygy equation~\cite{Larsen:2015ped,Georgoudis:2016wff},
\begin{equation}
b\frac{{\partial P_N^L}}{{\partial X}} + \sum\limits_a {{c_a}\frac{{\partial P_N^L}}{{\partial {x_a}}}}  = 0
\label{SZGY}
\end{equation}
with $b$ and $c_a$ being polynomials in $x_a$. 
 
Assuming that a solution of this equation has been found such that $b$ is independent of $x_a$ 
(eventually depending on external kinematics and internal masses and not identical to zero),
we have
\begin{eqnarray}
\frac{\partial }{{\partial X}}{F_{{\alpha _1}...{\alpha _N}}} &=& \left( {\frac{{ - d + E + 1}}{2}} \right)\frac{1}{G}\frac{{\partial G}}{{\partial X}}{F_{{\alpha _1}...{\alpha _N}}} \nonumber \\
&+&C_N^L{G^{\left( { - d + E + 1} \right)/2}}\int {\frac{{d{x_1}...d{x_N}}}{{x_1^{{\alpha _1}}...\,\,x_N^{{\alpha _N}}}}} \left( -{\sum\limits_a {\frac{{{c_a}}}{b}\frac{\partial }{{\partial {x_a}}}P{{_N^L}^{\left( {d - M - 1} \right)/2}}} } \right)
\end{eqnarray}
Then integrating by parts the second term in the rhs of the above equation and assuming that surface terms are vanishing (a standard assumption through Baikov representation) we get
\begin{eqnarray}
\frac{\partial }{{\partial X}}{F_{{\alpha _1}...{\alpha _N}}} &=& \left( {\frac{{ - d + E + 1}}{2}} \right)\frac{1}{G}\frac{{\partial G}}{{\partial X}}{F_{{\alpha _1}...{\alpha _N}}}  \\
&+&C_N^L{G^{\left( { - d + E + 1} \right)/2}}\int {d{x_1}...d{x_N}} {P_N^L}^{\left( {d - M - 1} \right)/2} \left\{ {\sum\limits_a {\frac{\partial }{{\partial {x_a}}}\left( {\frac{{{c_a}}}{b}\frac{1}{{x_1^{{\alpha _1}}...\,\,x_N^{{\alpha _N}}}}} \right)} } \right\}\nonumber 
\end{eqnarray}
The term in the curly bracket is easily seen to be a sum of terms of the form 
\[\frac{1}{{x_1^{{{\alpha '}_1}}...\,\,x_N^{{{\alpha '}_N}}}}\]
The powers $\alpha^\prime_a$ depend on the actual form of the solution of the syzygy equation, Eq.~(\ref{SZGY}). The result is as expected
\begin{equation}
\frac{\partial }{{\partial X}}{F_{{\alpha _1}...{\alpha _N}}} = \sum\limits_i {{R_i}}\,\, {F_{\alpha _1^{\left( i \right)}...\alpha _N^{\left( i \right)}}}
\label{eq:bde}
\end{equation}
with coefficients $R_i$ that are rational functions of the space-time dimension $d$, the external kinematics and the internal masses.
The rhs of the above equation contains integrals that are in general not MI.
We have verified in numerous examples, that after applying a standard IBP reduction to MI for the rhs of the above equation, the resulting differential equations for the MI are the same as those obtained with the standard approach.
It is still interesting to note that the initial form, Eq.~(\ref{eq:bde}), is generally not. 

\section{Cutting Feynman Integrals}
\label{sCUT}

Cutting FI in the Baikov representation has a very natural definition. Indeed we define an $n-$cut as follows
\begin{equation}
{F}_{{\alpha _1}...{\alpha _N}}|_{n \times \text{cut}} \equiv C_N^L{\left( G \right)^{\left( { - d + E + 1} \right)/2}}\left( {\prod\limits_{a = n+1}^{N} {\int {d{x_a}} } } \right)\left( {\prod\limits_{c = 1}^n {\oint\limits_{{x_{c = 0}}} {d{x_c}} } } \right)\frac{1}{{x_1^{{\alpha _1}} \ldots x_N^{{\alpha _N}}}} {P_N^L}^{\left( {d - M - 1} \right)/2}
\label{cut}
\end{equation}
where the Baikov variables $\{x_a:a=1,...,N\}$ have been divided in two subsets, containing $n$ cut propagators and $(N-n)$ uncut ones. The cut operation defined above is operational in any space-time dimension $d$
and for any FI given by Eq.~(\ref{FI}). Notice that the definition of the cut, Eq.~(\ref{cut}), is not identical to the traditional unitarity cut, see for instance Section 8.4 of ref.~\cite{Veltman:1994wz}, due to the lack of the
$\theta$-function constraint on the energy, and therefore it is not directly related to the discontinuity of the FI~\cite{Abreu:2014cla,Abreu:2015zaa}.

Let us now consider a set of MI, ${F_i} \equiv {F_{\alpha _1^{\left( i \right)}...\alpha _N^{\left( i \right)}}}$, $i=1,...,I$, satisfying a system of DE, with respect to variables $X_j$,
\begin{equation}
\frac{\partial }{{\partial {X_j}}}{F_i} = \sum\limits_{l = 1}^I {M_{il}^{\left( j \right)}{F_l}} 
\label{fullDE}
\end{equation}
with matrices $M^{(j)}$ depending on kinematical variables, internal masses, and the space-time dimension, $d$. Since the derivation of DE in Section~\ref{sDE} is insensitive to the cut operation, as defined in Eq.~(\ref{cut}), we may immediately write\footnote{Care should be taken in defining the DE so that no symmetries of MI are used that may be violated by the corresponding $n-$cut.} 
\begin{align}
\frac{\partial }{\partial X_j} F_i |_{n \times \text{cut}} = \sum_{l = 1}^I M_{il}^{(j)} F_l |_{n \times \text{cut}}
\label{cutDE}
\end{align}
with $F |_{n \times \text{cut}}$ representing an arbitrary $n-$cut: in other words, the cut integrals satisfy the same DE as the uncut ones\footnote{See also ref.~\cite{Anastasiou:2002yz,Lee:2012te,Larsen:2015ped} for related considerations.}. Of course for a given $n-$cut many of the MI
that are not supported on the corresponding cut vanish identically. Nevertheless, Eq.~(\ref{cutDE}) remains valid. Especially for the maximally cut integrals defined so that $n$ is equal to the number of
propagators (with $\alpha_i > 0$) of the integral, all integrals not supported on the cut vanish and the resulting DE is restricted to its homogeneous 
part. Evaluating the maximally cut MI provides therefore a solution to the homogeneous equation~\cite{Henn:2014qga,Primo:2016ebd}. Non-maximally cut integrals, on the other hand, can resolve non-homogenous parts
of the DE as well~\cite{Henn:2014qga}. 

One important implication is that cut and uncut integrals, although very different in many respects, as for instance their structure in $\epsilon-$expansion ($\epsilon\equiv (4-d)/2$), they are expressed in terms of the same class of functions\footnote{See also related discussion in ref.~\cite{CaronHuot:2012ab}, section 3.4.1.}. This 
is particularly important if we want to know {\it a priori} if a system of DE can be solved, for instance, in terms of Goncharov Polylogarithms, or if the solution contains a larger
class of functions including, for instance Elliptic Integrals. 

In Appendix \ref{app:examples} we have collected several results of maximally cut MI: in \ref{subs:double-box-m} we study a double-box with a massive loop, and find that it is expressible in terms of Polylogarithmic functions; 
in \ref{subs:sunset} we study two-loop sunset graph and show 
that only the fully massive one is elliptic\footnote{By elliptic we mean that it is expressible only in terms of Elliptic Integrals.}; 
in \ref{subs:bubble-triangle} we show how results can be obtained beyond $d=4$ and verified that certain maximally cut integrals as well as certain combinations of maximally cut integrals 
are expressed in terms of GPs in exactly the same way as their uncut counterparts;
in \ref{subs:box-triangle} we show that the cut of the elliptic box-triangle integral~\cite{Bonciani:2016qxi} is elliptic as well, 
and finally in \ref{subs:double-box-ell} we study the elliptic double-box~\cite{Bonciani:2016qxi}, and verify that the Elliptic Integrals only enter through its sub-topologies.

\section{Discussion and Outloook}
\label{Sdisc}

In this paper we have studied properties of Feynman integrals in Baikov representation. We have shown how to determine the limits of integration and how to obtain DE with respect 
to external kinematics and internal masses. We have introduced also a loop-by-loop approach in constructing the Baikov representation, so that for certain FI a smaller number
of integration variables is obtained. Then we provided a definition of a cut integral, operational in $d$ dimensions, and show that a cut integral satisfies the same system of DE as the 
uncut, original integral. We have shown how to compute the simplest, i.e. maximally, cut integral in Baikov representation, and give explicit results for several cases. 

Based on the fact  that cut integrals satisfy the same system of DE as the full, uncut integrals we have verified that
their analytic expressions are given in terms of the same class of functions, such as Goncharov Polylogarithms or Elliptic Integrals. 
We have therefore arrived at the conclusion that in a family of MI satisfying a given system of DE, the study of the maximally cut integrals for all its members 
can provide a {\it necessary and sufficient} criterion for the existence of a canonical form of the DE, and in the case when such a canonical form does not exist,
it provides solutions of the homogeneous
parts of the system of DE (see also ref.~\cite{Henn:2014qga,Primo:2016ebd}). An application of these ideas to non-planar pentabox integrals will be discussed elsewhere.

Baikov representation is well suited for these considerations, drastically simplifying the computation of cut integrals for arbitrary external momenta and internal masses. 
It is still an open question if it can also be used to actually compute the MI. To this end,
an algorithm,
allowing the resolution of singularities in $\epsilon$, 
needs to be devised. 
It remains to be seen
if this is possible and more importantly what kind of integral representations for the individual terms
in this expansion such an algorithm produces.

\subsection*{Acknowledgements}
We gratefully acknowledge fruitful discussions with P.~A.~Baikov, V.~A.~Smirnov,  M.~Czakon, R.~N.~Lee, E.~Panzer, C.~Anastasiou, K.~J.~Larsen, J.~Henn and C.~Duhr during various stages of this project.
This research was supported through the Initial Training Network HiggsTools under contract
PITN-GA-2012-316704.

\appendix
\section{Alternative derivation of the Baikov representation}
\label{app:derivation}

In this appendix we will show an alternative derivation of the Baikov representation of Feynman integrals, complementary to the one given in section \ref{sBaikov} in the main text.

We start from Eq. (\ref{FI}) for the one-loop case
\begin{align}
F_{\alpha_1 \ldots \alpha_N} = \int \! \frac{d^d k}{i \pi^{d/2}} \, \frac{1}{D_1^{\alpha_1} \cdots D_N^{\alpha_N}}
\label{eq:FIapp}
\end{align}

The integrand $( D_1^{\alpha_1} \cdots D_N^{\alpha_N} )^{-1}$ depends on $E$ independent external momenta, which allows us to split the integration into an $E$-dimensional ``parallel'' subspace $k_{||}$ and a $(d-E)$-dimensional ``orthogonal'' subspace $k_{\perp}$. Notice that the integrand depends on the orthogonal directions only through  $k_{\perp}^2$, allowing us to perform the angular part of the integral over the orthogonal space
\begin{align}
F_{\alpha_1 \ldots \alpha_N} = \frac{\pi^{-E/2}}{i \, \Gamma((d-E)/2)} \int \! \frac{\lambda^{(d-E-2)/2}}{D_1^{\alpha_1} \cdots D_N^{\alpha_N}} \, d^E k_{||} d \lambda
\label{eq:FIinterm}
\end{align}
where we define $\lambda \define k_{\perp}^2$.


We change now to a different set of variables $\varsigma_i \define k \cdot p_i$ which is equivalent to $s_{1,E+i}$ from section \ref{sBaikov}. Defining $G$ as the Gram determinant of the external momenta, it is not hard to realize that $d^E k_{||} = - (-G)^{-1/2} d^E \varsigma$, and we now want to change the integration to the set of Baikov variables $x$. 
Using the same argument as in section \ref{sBaikov}, we get that $d^E k_{||} d \lambda = \det(A^{-1}) d^{E+1} x$ where $A$ is the matrix defined in Eq.~(\ref{eq:x-def}),
and we see that $\det(A^{-1}) = \pm 2^{-E}$ depending on the sign of the external momenta in the definition of the loop momenta. Putting this together, yields
\begin{align}
F_{\alpha_1 \ldots \alpha_N} = \frac{i \pi^{-E/2}}{\Gamma((d-E)/2)} \frac{\det(A^{-1})}{\sqrt{-G}} \int \! \frac{\lambda^{(d-E-2)/2}}{x_1^{\alpha_1} \cdots x_N^{\alpha_N}} \, d^{E+1} x
\label{eq:FIbai}
\end{align}

If additionally we use that $\lambda = P/G$ where $P$ -- the Baikov polynomial -- is given as the Gram determinant of the full set of momenta $\{q\} = \{k,p_1,\ldots,p_E\}$, we see that Eq. \eqref{eq:FIbai} is equivalent to Eq. \eqref{BR} in the one-loop case.
The above derivation can be straightforwardly generalised for any numbers of loops.

\subsection*{The loop-by-loop case}

We may now go through the same procedure as above, loop by loop,  for a two-loop process. A traditional Baikov representation of a two-loop Feynman integral with $E$ independent external momenta would involve $3+2E$ integrations, but in most cases it is possible to get the number further down. Each individual loop will in general be dependent only on a subset of the $E$ external momenta. Let us denote with $k_2$ the loop momentum of the loop with the smallest such subset, and the size of that subset $E_2-1$. This means that the integrand of this loop depends on $E_2$ different momenta including $k_1$. Performing the transverse angular part of the $k_2$ integration individually with the method leading up to Eq. \eqref{eq:FIinterm}, gives
\begin{align}
F^{\text{two-loop}}_{\alpha_1 \ldots \alpha_N} = \frac{- \pi^{-(E_2+d)/2}}{\Gamma((d-E_2)/2)} \int \! \frac{\lambda_{22}^{(d-E_2-2)/2}}{D_1^{\alpha_1} \cdots D_N^{\alpha_N}} \;\; d^{E_2} {k_2}_{||} d \lambda_{22} d^d k_1
\end{align}
where $\lambda_{22} = {k_2}_{\perp}^2$.

The integrand now depends on all the external momenta, so the transverse angular part of the $k_1$ integration may be done, leaving $E+E_2+2$ integrations in total. 
So only in the cases where both loops depend on $E$ external momenta (so $E_2 = E+1$)  the number of integrations will be equal to that of the standard Baikov approach, otherwise it will be smaller.

Changing integration variables from $k_{||}$ to $\varsigma$ and from $\varsigma$ and $\lambda$ to the Baikov variables $x$, our result for the two-loop case in the loop-by-loop approach
is written as
\begin{align}
F^{\text{two-loop}}_{\alpha_1 \ldots \alpha_N} = \frac{- \pi^{-(E+E_2)/2}}{\Gamma((d-E)/2) \, \Gamma((d-E_2)/2)} \frac{\det(A^{-1})}{\sqrt{-G_1}} \int \! \frac{\lambda_{11}^{(d-E-2)/2} \lambda_{22}^{(d-E_2-2)/2}}{\sqrt{-G_2} \; 
x_1^{\alpha_1} \cdots x_N^{\alpha_N}} \, d^{E+E_2+2} x
\label{eq:FIbailbl}
\end{align}
with $G_1$ the Gram determinant of the $E$ external momenta and $G_2$ the Gram determinant of the $E_2$ different momenta including $k_1$.

\section{Examples of cuts performed in Baikov variables}
\label{app:examples}

In the appendix we will go through several examples of maximally cut Feynman integrals.

\subsection{Double-box with massive loop}
\label{subs:double-box-m}

As a first example we will do a double-box with one massive loop. That integral is given as
\begin{align}
F_{\text{double-box}} &= \int \! \frac{d^d k_1}{i \pi^{d/2}}  \frac{d^d k_2}{i \pi^{d/2}} \frac{1}{x_1 x_2 x_3 x_4 x_5 x_6 x_7}
\end{align}
with
\begin{align}
x_1 &= k_1^2-m^2 & x_2 &= (k_1+p_1)^2-m^2 \quad & x_3 &= (k_1+p_1+p_2)^2-m^2 \nn \\
x_4 &= (k_2+p_1+p_2)^2 \quad & x_5 &= (k_2-p_4)^2 & x_6 &= k_2^2 \nn \\
x_7 &= (k_1-k_2)^2-m^2 & x_8 &= (k_1-p_4)^2-m^2
\end{align}
where $x_8$ is needed for later. Using the loop-by-loop approach, Eq.~(\ref{eq:FIbailbl}), we may perform the angular parts of the integral, leaving integrals over the eight Baikov variables
\begin{align}
F_{\text{double-box}} = \frac{- \pi^{-3}}{\Gamma^2((d-3)/2)} \frac{\det(A^{-1})}{\sqrt{-G_1}} \int \! \frac{\lambda_{11}^{(d-5)/2} \lambda_{22}^{(d-5)/2}}{\sqrt{-G_2} \; x_1 \cdots x_7} \, d^{8} x
\label{eq:FIbailbl}
\end{align}
The various quantities are\footnote{This is the only example in which these quantities will be written out in full}
\begin{align}
\lambda_{11} &= \bigg( 4 m^2 s t (s+t) + 2 s t \Big( 2 x_1 x_3 - x_1 x_2 - x_2 x_3 + t (x_1+x_3) - (x_1-2 x_2+x_3) x_8 \Big) \nn \\
& \;\;\; - s^2 \Big( t^2+(x_2-x_8)^2-2 t (x_2+x_8) \Big) - t^2 (x_1-x_3)^2 \bigg)/ \Big(4 s t (s+t) \Big)
\end{align}
\begin{align}
\lambda_{22} &= -\bigg( 4 m^2 s \Big( -(s x_5)+(x_4-x_5) (x_5-x_6) \Big) + \Big( x_3 (-x_5+x_6)+s (x_5-x_7) \Big)^2 \nn \\
&\;\; + 2 \Big(x_3 (x_4-x_6) (-x_5+x_6)-s^2 (x_5+x_7)+s (x_4 x_5-2 x_4 x_6+x_5 x_6+x_3 (x_5+x_6) \nn \\
&\;\; + (x_4-2 x_5+x_6) x_7) \Big) x_8 + \Big( s^2+(x_4-x_6)^2-2 s (x_4+x_6) \Big) x_8^2 \nn \\
&\;\; + x_1^2 (x_4-x_5)^2 + 2 x_1 \Big(s (-2 x_3 x_5+(x_4-x_5) (x_5-x_7)+(x_4+x_5) x_8) \\
&\;\; +(x_4-x_5) (x_3 (x_5-x_6)+(-x_4+x_6) x_8) \Big) \bigg)/ \Big( 4 s (m^2 s+(x_1-x_8) (x_3-x_8)+s x_8) \Big) \nn
\end{align}
\begin{align}
& G_1 = -st(s+t)/4 \nn \\
& G_2 = -s \Big( m^2 s+x_1 (x_3-x_8)+x_8 (s-x_3+x_8) \Big)/4  \\
& \det \big( A^{-1} \big) = 1/64 \nn
\end{align}
and $s = {\left( {{p_1} + {p_2}} \right)^2},t = {\left( {{p_1} + {p_4}} \right)^2},s > 0,t < 0,s+t>0,{m^2} > 0$.

Inserting all this, cutting the seven propagators, expanding in $\epsilon$, and renaming $x_8$ to $z$, gives
\begin{align}
F_{\text{double-box}|7 \times \text{cut}} &= \frac{1}{4 \pi^4} \frac{1}{s} \, \int_{r_-}^{r_+} \! dz\frac{1}{z \sqrt{P(z)}} \; + \; \kur{O}( \epsilon)
\end{align}
where
\begin{align}
P(z) &= -4 m^2 s^2 t-4 m^2 s t^2+s^2 t^2-2 s^2 t z+s^2 z^2
\end{align}
and $r_{\mp}$ are the two roots of $P(z)$.
The result is
\begin{align}
F_{\text{double-box}|7 \times \text{cut}} &= \frac{1}{4 \pi^4} \; \frac{i\pi}{s \sqrt{s t ( s t - 4 m^2 (s+t))}} \; + \; \kur{O}( \epsilon).
\end{align}
In fact we have computed also the order $\epsilon$ which is given 
in terms of weight 2 functions, but since it is quite extensive we give the result for another MI, defined as follows,
\begin{align}
F_{\text{double-box-N}} &= \int \! \frac{d^d k_1}{i \pi^{d/2}}  \frac{d^d k_2}{i \pi^{d/2}} \frac{x_8}{x_1 x_2 x_3 x_4 x_5 x_6 x_7}
\end{align}
whose expression is simpler,
\begin{equation}
I\equiv- 4i{\pi ^3}{e^{2{\gamma _E}\epsilon }}{s^{2 + \epsilon }}{\left( { - t} \right)^{ - \epsilon }}{\left( {s + t} \right)^{ - \epsilon }}{\left( {{m^2}} \right)^{3\epsilon }}F_{\text{double-box-N}|7 \times \text{cut}} 
\end{equation}
\begin{eqnarray}
I&=& 1 - \epsilon \log \left( {\frac{{ - 4{t^3}{{\left( {s + t} \right)}^2}{{\left( {\sqrt {{Y_ + }}  + \sqrt {{Y_ - }} } \right)}^4}}}{{{m^2}{{\left( {\sqrt {{X_{ +  - }}}  + \sqrt {{X_{ -  - }}} } \right)}^2}{{\left( {\sqrt {{X_{ -  + }}}  + \sqrt {{X_{ +  + }}} } \right)}^2}}}} \right)+\kur{O}( \epsilon^2)\nn\\
{X_{ \pm  \pm }} &=& ms\sqrt {t(s - 4{m^2})(s + t)}  \pm 4{m^2}t(s + t) \pm m\sqrt {st\left( {s + t} \right)} (s + 2t)\\
{Y_ \pm } &=& 2m\sqrt {s + t}  \pm \sqrt {st} \nn
\end{eqnarray}
suggesting that the solution is expressible in terms of logarithmic/polylogarithmic functions.

\subsection{The sunset}
\label{subs:sunset}

We will here go through the same considerations for the well-studied sunset-integral~\cite{Adams:2015gva}. 
The integral is given as
\begin{align}
F_{\text{sunset}} &= \int \! \frac{d^d k_1 \, d^d k_2}{\big(i \pi^{d/2}\big)^2} \frac{1}{x_1 x_2 x_3}
\end{align}
with
\begin{align}
x_1 &= (k_1-p)^2 - m_1^2  &  x_2 &= k_2^2 - m_2^2  &  x_3 &= (k_1-k_2)^2 - m_3^2  &  x_4 &= k_1^2.
\end{align}
Using the loop-by-loop approach, we get
\begin{align}
F_{\text{sunset}} &= \frac{- \pi^{-1}}{\Gamma^2((d-1)/2)} \frac{\det(A^{-1})}{\sqrt{-G_1}} \int \! \frac{\lambda_{11}^{(d-3)/2} \lambda_{22}^{(d-3)/2}}{\sqrt{-G_2} \; x_1 x_2 x_3} \, d^4 x
\label{eq:sunsetBaikovLBL}
\end{align}
with
\begin{align}
\lambda_{22}=\lambda(x_4,m_2^2,m_3^2)\;\;\;\;\;\;\lambda_{11}=\lambda(s,x_4,m_1^2)
\end{align}
$\lambda(s_1,s_2,s_3)\equiv s_1^2+s_2^2+s_3^2-2s_1 s_2-2s_2 s_3-2 s_3 s_1$, is the K\"all\'en function and $s=p^2$.
Performing the triple cut gives~\cite{Remiddi:2016gno},
\begin{align}
F_{\text{sunset}| 3 \times \text{cut}} &\prop   s^{-1+\epsilon} \int dx_4 x_4^{-1+\epsilon} \left(\lambda_{11}\lambda_{22}\right)^{1/2-\epsilon}
\end{align}
Expanding in $\epsilon$  
\begin{align}
F_{\text{sunset}| 3 \times \text{cut}} &\prop   s^{-1} \int dx_4 x_4^{-1} \left(\lambda_{11}\lambda_{22}\right)^{1/2}+\kur{O}(\eps)
\label{eq:sunsetelliptic}
\end{align}
it is easily seen that this is an integral over the square-root of a quartic polynomial~\cite{Remiddi:2016gno}, 
which yields elliptic integrals, and such functions are indeed present in the result for the full sunset integral as well.
On the other hand, if any of the masses vanish, the square root partially factorises, and the result is  expressible in terms of polylogarithmic functions.

Notice, that the extra cut over $x_4$ ($m_2 > m_3$), after expanding in $\epsilon$,
\begin{align}
\oint\limits_{{x_4} = 0} {}  dx_4 x_4^{-1} \left(\lambda_{11}\lambda_{22}\right)^{1/2} & = (s-m_1^2)(m_2^2-m_3^2)
\label{eq:extracut}
\end{align}
yields a rational term at order $\kur{O}(\eps^0)$, which seems to contradict our previous findings, Eq.~(\ref{eq:sunsetelliptic}), as well as the known result for the full integral~\cite{Adams:2015gva}.
Nevertheless, it is easy to verify, starting from Eq.~(\ref{eq:sunsetBaikovLBL}), that $F_{\text{sunset}| 4 \times \text{cut}} =0$ and that the result of Eq.~(\ref{eq:extracut}) is just an artefact of expanding in $\epsilon$ before
cutting.

\subsection{The bubble-triangle}
\label{subs:bubble-triangle}


As mentioned in Section~\ref{sIntro}, the introduction of the idea of canonical bases~\cite{Henn:2013pwa} caused a renaissance in the derivation of analytical expressions of Feynman integrals. We will not go through the details of what a canonical basis implies, merely say that  no general algorithm for obtaining such a basis exists. In ref.~\cite{Henn:2014qga} it was implied that canonical integrals have maximal cuts which equal a number (rather than a function of kinematical variables). This criterion significantly limits the search space, and has been a helpful guideline in many cases.

Yet for some cases the maximal cut derived in the traditional way in  four dimensions, does not provide the full information obtainable from such a criterion. For an example lets us re-examine the triangle with bubble insertion discussed in section \ref{sLBL}. We will here parametrize it as
\begin{align}
F_{\alpha_1 \alpha_2 \alpha_3 \alpha_4 \alpha_5} &= \int \! \frac{d^d k_1}{i \pi^{d/2}}  \frac{d^d k_2}{i \pi^{d/2}} \frac{x_5^{- \alpha_5}}{x_1^{\alpha_1} x_2^{\alpha_2} x_3^{\alpha_3} x_4^{\alpha_4}}
\end{align}
with
\begin{align}
x_1 &= (k_1-p_1)^2 & x_2 &= (k_1+p_2)^2 & x_3 &= k_2^2 & x_4 &= (k_1-k_2)^2 & x_5 &= k_1^2
\end{align}
Using the loop-by-loop approach of eq. \eqref{eq:FIbailbl}, we may turn this into a five-fold integral
\begin{align}
F_{\alpha_1 \alpha_2 \alpha_3 \alpha_4 \alpha_5} = \frac{- \pi^{-3/2}}{\Gamma((d-2)/2) \, \Gamma((d-1)/2)} \frac{\det(A^{-1})}{\sqrt{-G_1}} \int \! \frac{\lambda_{11}^{(d-4)/2} \lambda_{22}^{(d-3)/2} \!\!\!\!\!\!}{\sqrt{-G_2}} \; \frac{x_5^{-\alpha_5} \; d^{5} x}{x_1^{\alpha_1} x_2^{\alpha_2} x_3^{\alpha_3} x_4^{\alpha_4}}
\end{align}

The set of canonical differential equations used in ref.~\cite{Henn:2014lfa} contains two integrals with the topology of this integral, and they are given as ($s = {\left( {{p_1} + {p_2}} \right)^2}$)
\begin{align}
I_1 &= \epsilon R_{12} F_{11210} & I_2 &= \Big( s F_{1221 - \! 1} - \tfrac{1}{2} \epsilon \big( p_1^2 - p_2^2 - s \big) F_{11210} \Big)
\label{eq:I12def}
\end{align}
where $R_{12}$ denotes the K{\"a}ll{\'e}n function
\begin{align}
R_{12} &= \sqrt{p_1^2 + (p_2^2-s)^2 - 2 p_1^2 (p_2^2+s)}
\end{align}
Performing the cut in four dimensions will only catch the leading term in $\epsilon$, and thus not the second term in $I_2$. 
On the contrary, the definition of maximal cut given in Eq.~(\ref{cut}), allows us to compute the cut integral in $d$ dimensions and for arbitrary powers of propagators in the integrand. 
The result is given by 
\begin{align}
F_{11210|4 \times \text{cut}} &\prop 2 (d-3) \, R_{12}^{(3-d)/2} \\
& \;\;\; \times \int_{r_-}^{r_+} z^{(d-6)/2} \, \big( s ( p_1^2 p_2^2 - p_1^2 z - p_2^2 z + s z + z^2 ) \big)^{(d-4)/2} \; \id z \nn \\
F_{1221 - \! 1|4 \times \text{cut}} &\prop (d-3) (d-4) \, R_{12}^{(3-d)/2} \int_{r_-}^{r_+} z^{(d-4)/2} \nn \\
& \;\;\; \times \big( -p_1^2 p_2^2 + p_2^4 - p_2^2 s + p_1^2 z - p_2^2 z - s z \big) \\
& \;\;\; \times \big( s ( p_1^2 p_2^2 - p_1^2 z - p_2^2 z + s z + z^2 ) \big)^{(d-6)/2} \; \id z \nn
\end{align}
where $z = x_5$ and $r_\pm$ the roots of the polynomial: $p_1^2 p_2^2 - p_1^2 z - p_2^2 z + s z + z^2 $.
These integrals evaluate to hypergeometric functions, and as the second integral diverges in the $d \rightarrow 4$ limit, the integration does not commute with the $\epsilon$ expansion. 
Thus one has to evaluate them in $d$ dimensions and then expand the resulting functions, for instance using the HypExp package~\cite{Huber:2007dx}, to the desired order in $\epsilon$. 
The result for the integrals defined in Eq.~(\ref{eq:I12def}) is given by 

\begin{eqnarray}
I_{1\,|4 \times \text{cut}}&=&
\frac{ 2^{4 \epsilon -3}\epsilon  \cos (\pi  \epsilon ) \Gamma \left(\epsilon +\frac{1}{2}\right) }{\pi ^2 \Gamma \left(\frac{3}{2}-\epsilon \right)}
{\left(p_1^2\right)}^{-2 \epsilon } x^{-\epsilon } (x+1)^{-\epsilon }(y-1)(x y+1)^{-\epsilon }
\nn\\
&\times&\, _2F_1(1-\epsilon ,\epsilon +1;2-2 \epsilon ;1-y) 
\end{eqnarray}
\begin{equation}
I_{2\,|4 \times \text{cut}}=
\frac{4^{2 \epsilon -1} }{\pi  \Gamma \left(\frac{1}{2}-\epsilon \right)^2}
{\left(p_1^2\right)}^{-2 \epsilon } x^{-\epsilon } (x+1)^{-\epsilon } (x y+1)^{-\epsilon }\, _2F_1(-\epsilon ,\epsilon ;-2 \epsilon ;1-y) 
\end{equation}
where the 
variables $x$,$y$ have been introduced, defined through $s \equiv p_1^2 (1+x) (1+xy)$ and $p_2^2 \equiv p_1^2 x^2 y$.

In order to explicitly show the uniform transcendentallity  property of these functions, we present the results expanded up to order $\eps^3$,
\begin{eqnarray}
N_\epsilon I_{1\,|4 \times \text{cut}} &=&
\epsilon  \log (y)+\epsilon ^2 \left(-2\, \text{Li}_2(1-y)-\log ^2(y)\right)+\epsilon ^3 \Big( -4\, \text{Li}_3(1-y)-2\, \text{Li}_3(y)
\nn\\
&-&\left. \text{Li}_2(y) \log (y)+\frac{2}{3} (\log (y)-3 \log (1-y)) \log ^2(y)+2\, \zeta (3)\right)+ \kur{O}(\eps^4)
\end{eqnarray}
\begin{eqnarray}
N_\epsilon I_{2\,|4 \times \text{cut}} &=& 
1-\frac{1}{2} \epsilon  \log (y)+\frac{1}{2} \epsilon ^2 \left(\log ^2(y)-\pi ^2\right)+ \frac{1}{12} \epsilon ^3 \Big( 36\, \text{Li}_3(y)+18\, \text{Li}_2(1-y) \log (y)
\nn\\
&-& 4 \log ^3(y)+18 \log (1-y) \log ^2(y)-3 \pi ^2 \log (y)-92\, \zeta (3) \Big)+ \kur{O}(\eps^4)
\end{eqnarray}
with $N_\epsilon=e^{2 \gamma_E  \epsilon }(p_1^2)^\epsilon x^{\epsilon} (x+1)^{\epsilon} (x y+1)^{\epsilon}$. 

\subsection{The elliptic box-triangle}
\label{subs:box-triangle}

Here we will go through the generalized cut of the elliptic box-triangle studied in ref. \cite{Bonciani:2016qxi} where it is denoted $I^A_{66}$, and in ref. \cite{Primo:2016ebd}. It is given as
\begin{align}
F_{\text{box-triangle}} = \int \! \frac{d^d k_1 \, d^d k_2}{\big(i \pi^{d/2} \big)^2} \, \frac{1}{x_1 x_2 x_3 x_4 x_5 x_6}
\end{align}
with
\begin{align}
x_1 &= k_1^2-m^2 & x_2 &= (k_1+p_1)^2-m^2 \quad & x_3 &= (k_1+p_1+p_2)^2-m^2 \nn \\
x_4 &= (k_2-p_4)^2-m^2 \quad & x_5 &= k_2^2-m^2 & x_6 &= (k_1-k_2)^2 \nn \\
x_7 &= (k_1-p_4)^2
\end{align}
where $x_7$ has been introduced for later convenience. The kinematics is such that
\begin{align}
p_1^2 = p_2^2 &= 0 & (p_1+p_2)^2 &= s & (p_1+p_4)^2 &= t & (p_2+p_4)^2 &= u = p_4^2-s-t
\end{align}

With the loop-by-loop approach as described in Section \ref{sLBL}, we see that the $k_2$ integral has a transverse space with dimension $2$, as it is formed by $p_4$ and $k_1$, while the $k_1$ integral has a transverse space formed by $p_1$, $p_2$, and $p_4$, which has dimension $3$. Thus we can write $F_{\text{box-triangle}}$ as a $7$-dimensional integral over the seven Baikov variables given above.
\begin{align}
F_{\text{box-triangle}} &= \frac{- \, \pi^{-5/2}}{\Gamma(\tfrac{d-2}{2}) \Gamma(\tfrac{d-3}{2})} \, \frac{\det(A^{-1})}{\sqrt{-G_1}} \int \frac{1}{x_1 \cdots x_6} \; \frac{\lambda_{22}^{(d-4)/2} \, \lambda_{11}^{(d-5)/2}}{\sqrt{-G_2}} \, \id^{7} x
\end{align}
Here $\lambda_{22}$ is a function of all the Baikov variables, and $\lambda_{11}$ and $G_2$ of those not containing $k_2$. We may then do the hexa-cut of the integral, yielding an integral over one variable $z=x_7$. That integral is well-behaved in the $d=4$ limit, and thus the integration and the $\epsilon$ expansion commute, giving
\begin{align}
F_{\text{box-triangle}|6 \times \text{cut}} &= C \int_{r_-}^{r_+} \! \frac{\id z}{\sqrt{F_1(z) F_2(z)}} \, + \kur{O}(\epsilon)
\end{align}
where $C = i/(4 \pi^3)$ and
\begin{align}
F_1(z) &= m^4 - 2 m^2 p_4^2 + p_4^4 - 2 m^2 z - 2 p_4^2 z + z^2 \nn \\
F_2(z) &= s \big( m^4 s + 2 m^2 (2 t u + s (t - z)) + s (t - z)^2 \big)
\end{align}
and $r_\mp$ are the two roots of $F_2(z)$ which are given as
\begin{align}
r_{\mp} &= (m^2+t) \mp \sqrt{-m^2 stu}/s
\end{align}

Following~\cite{Byrd:1971}, we may perform the integral with the result
\begin{align}
F_{\text{box-triangle}|6 \times \text{cut}} &= \frac{2 i C}{\sqrt{X}} \, K \! \left( \frac{-16 m^2 \sqrt{- p_4^2 stu}}{X} \right)+ \kur{O}(\epsilon)
\end{align}
where $K(k^2)$ is the complete elliptic integral of the first kind, and where
\begin{align}
X &= s (p_4^2 - t)^2 - 4 m^2 \left(p_4^2 s - t u + 2 \sqrt{-p_4^2 s t u}\right)
\end{align}
which holds in the physical region with $m^2>0$, $s>0$, $p_4^2>0$, $t<0$, $u<0$.

\subsection{The elliptic double-box}
\label{subs:double-box-ell}

Here we will go through the generalized cut of the elliptic double-box studied in ref.~\cite{Bonciani:2016qxi} where it is denoted $I^A_{70}$. It is given as
\begin{align}
F_{\text{ell. double-box}} = \int \! \frac{d^d k_1 \, d^d k_2}{\big(i \pi^{d/2} \big)^2} \, \frac{1}{x_1 x_2 x_3 x_4 x_5 x_6 x_7}
\end{align}
with
\begin{align}
x_1 &= k_1^2-m^2 & x_2 &= (k_1+p_1)^2-m^2 \quad & x_3 &= (k_1+p_1+p_2)^2-m^2 \nn \\
x_4 &= (k_2+p_1+p_2)^2-m^2 \quad & x_5 &= (k_2-p_4)^2-m^2 & x_6 &= k_2^2-m^2 \nn \\
x_7 &= (k_1-k_2)^2 & x_8 &= (k_1-p_4)^2
\end{align}
The kinematics is as in the previous example.

Here each of the two integrals over the loop momenta $k_1$ and $k_2$ are over transverse spaces with dimensionality three, giving an eight-dimensional integral after the angular components have been integrated out:
\begin{align}
F_{\text{ell. double-box}} &= \frac{- \, \pi^{-3}}{\Gamma^2 \, (\tfrac{d-3}{2})} \, \frac{\det(A^{-1})}{\sqrt{-G_1}} \int \frac{1}{x_1 \cdots x_7} \; \frac{\lambda_{22}^{(d-5)/2} \, \lambda_{11}^{(d-5)/2}}{\sqrt{-G_2}} \, \id^{8} x
\end{align}

As before we may perform a cut of all the propagators -- an hepta-cut in this case, yielding an integral over the last variable variable $z=x_8$
\begin{align}
F_{\text{ell. double-box}|7 \times \text{cut}} &= \frac{C}{\sqrt{s(s-4 m^2)}} \int_{r_-}^{r_+} \! \frac{\id z}{z \sqrt{f(z)}} \, + \kur{O}(\epsilon)
\label{eq:elldb}
\end{align}
where $C = 1/(4 \pi^4)$ and
\begin{align}
f(z) &= s \big( 4 m^2 t u + s (t - z)^2 \big)
\end{align}
and $r_\mp$ are the two roots of $f(z)$ which are given as
\begin{align}
r_{\mp} &= t \mp 2 \sqrt{- m^2 stu}/s
\end{align}

That integral may be done and yields
\begin{align}
F_{\text{ell. double-box}|7 \times \text{cut}} &= \frac{-i}{4 \pi^3} \frac{1}{s \sqrt{(4 m^2 - s) t (st + 4 m^2 u)}} + \kur{O}(\epsilon)
\end{align}
verifying that the elliptic character of the uncut integral stems from the non-homogeneous part of the corresponding differential equation through its elliptic sub-topologies~\cite{Bonciani:2016qxi} .


\bibliographystyle{JHEP}
\bibliography{biblio}

\end{document}